\documentclass[USenglish,twocolumn]{article}

\usepackage{etex}
\usepackage[utf8]{inputenc}
\usepackage[big]{dgruyter_NEW}
\usepackage{latexsym}

\usepackage{enumerate}\usepackage{colortbl}

\usepackage{inputenc,epstopdf,subfig}
\inputencoding{utf8}

\usepackage{slashbox}

\usepackage{amsfonts,hypcap}
\usepackage[T1]{fontenc}
 \usepackage{multirow}
\usepackage[table]{xcolor}
\usepackage{rotating}
\usepackage{lscape}
\usepackage{longtable}
\usepackage{multicol}

\definecolor{light-gray}{gray}{0.95}

\newcommand*{\authorimg}[1]{%
  \raisebox{-.14\baselineskip}{%
    \includegraphics[
      height=5mm,
      keepaspectratio,
    ]{#1}%
  }%
}

\usepackage[absolute]{textpos}
\usepackage[resetfonts]{cmap}

\usepackage{textcomp}

\usepackage{appendix}
\makeatletter
\def\maketag@@@#1{\hbox{\m@th\normalfont\normalsize#1}}
\makeatother 

\usepackage{enumitem,tikz }
\usetikzlibrary{arrows}
\setlist[itemize]{topsep=5pt,leftmargin=25pt}
\setlist[enumerate]{topsep=5pt, leftmargin= 25pt}

\usepackage{acro}

\immediate\write18{makeindex \jobname.nlo -s nomencl.ist -o \jobname.nls}
\usepackage{nomencl}
\makenomenclature
\makeindex

\DeclareSymbolFont{arrows}{OMS}{cmsy}{m}{n}
\DeclareMathSymbol{\leftrightarrow}{\mathrel}{arrows}{"24}
\DeclareMathSymbol{\leftarrow}{\mathrel}{arrows}{"20}
   
\DeclareMathSymbol{\rightarrow}{\mathrel}{arrows}{"21}
   
\DeclareMathSymbol{\mapstochar}{\mathrel}{arrows}{"37}
\DeclareMathSymbol{\relbardash}{\mathbin}{arrows}{"00}

\acsetup{first-style=short}

\newcommand{\aj}{AJ}
\newcommand{\apj}{ApJ}
\newcommand{\apjl}{ApJ}
\newcommand{\apjs}{ApJS}
\newcommand{\aap}{A\&A}
\newcommand{\prd}{Phys.Rev.D.}

\DeclareMathSymbol{\prime}{\mathord}{symbols}{"30}
\setcitestyle{aysep={}}

\allowdisplaybreaks
\begin{document}
\begin{textblock}{165}(213,38)
\includegraphics[width=0.35cm]{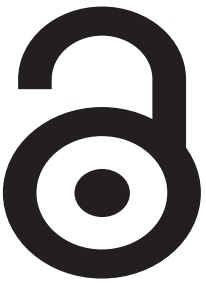}
\end{textblock}
\begin{textblock}{165}(43,283)
\noindent \sffamily\scriptsize \includegraphics[width=0.17cm]{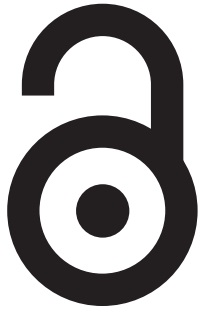} Open Access. \sffamily\scriptsize© 2017   D. F. Crawford \textit{et al.},  published by De Gruyter Open. \authorimg{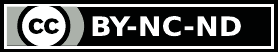} This work is licensed under the Creative Commons Attribution-NonCommercial-NoDerivatives 4.0 License
\end{textblock}

\articletype{Research Article{\hfill}}
\title{A problem with the analysis of  type Ia supernovae}

\runningtitle{Analysis of type Ia supernovae}

\author*[]{David F. Crawford}

\runningauthor{D. F. Crawford \textit{et al.}}

\received{Aug 16, 2017}
\accepted{Nov 16, 2017}

 \journalname{Open Astron.}
  \journalyear{2017}
  \journalvolume{26}
 \startpage{111}
   \DOI{\url{https://doi.org/10.1515/astro-2017-0013}}

\begin{abstract}
{Type Ia supernovae have light curves that have widths and magnitudes that can be used for testing cosmologies and they provide one of the few direct measurements of time dilation.
It is shown that  the standard analysis that calibrates the light curve against a rest-frame average (such as SALT2) removes all the cosmological information from the calibrated light curves. Consequently type Ia supernovae calibrated with these methods cannot be used to investigate cosmology.
The major evidence that supports the hypothesis of a static universe is that the  measurements of the widths of the raw light curves of type Ia supernovae do not show any time dilation.  The intrinsic wavelength dependence shown by the SALT2 calibration templates is also consistent with no time dilation.
Using a static cosmological model the peak absolute magnitudes of raw type Ia supernovae observations are also independent of redshift.  These results support the hypothesis of a static universe.}
\end{abstract}

\maketitle





{ \let\thempfn\relax
\footnotetext{\hspace{-1ex}{\Authfont\small \textbf{Corresponding Author: David F. Crawford:}} {\Affilfont Astronomical Society of Australia
44 Market St Naremburn 2065 NSW Australia,\newline
E-mail: dcrawfrd@bigpond.net.au,
Orcid: 0000-0003-0710-1683}}
}

\newcolumntype{Y}{>{\centering\arraybackslash}X}
\newcolumntype{P}{>{\centering\arraybackslash}p}

 \renewcommand{\figurename}{Figure}

\section{Introduction}
\label{s1}
Type Ia supernovae are transient phenomena that take about  twenty days to reach a peak brightness and then the brightness  decreases at a slower rate. Type Ia supernovae (for brevity SNe) are also known for their remarkably similar light curves  and this property  makes them excellent cosmological probes. For example the width of the light curves is one of the few cosmological observations that can directly measure time dilation at large redshifts.

The basis of the standard cosmological model is that the observed Hubble redshift of distant galaxies is  a consequence of an expanding universe. In this model redshift is consequence of an expansion of the space between galaxies and although it is mathematically the same as if it was produced by a velocity it is not a true velocity but a property of the distance between the galaxy and the observer. For example many different observers scattered around the universe would observe vastly different values of the redshift for the same observed galaxy. Then just like the Doppler effect any time interval must be dilated as a direct function of the pseudo velocity. Thus time dilation is an essential property of any expanding universe.

The observed Hubble redshift, $z$,  is defined as the ratio of the observed wavelength to the emitted wavelength, minus one. In an expansion model the ratio of any observed time  period to the emitted time period is identical to the ratio of the wavelengths, namely $(1+z)$. This is true for any time interval and is the time dilation that is applicable to the widths of the supernova  light curves. Clearly the equivalent pseudo velocity is $cz$.

Any challenge to the standard model must show that observations of SNe light-curve widths do not have time dilation even though the observed spectral lines  show the Hubble redshift. An essential requirement of this analysis is that  the  wavelength of photons is inversely proportional to their energy. Thus a photon that has lost energy since its emission will be observed to have a lower energy and the observed Hubble redshift may be due to some "tired light" process like that described  in \citep{Crawford09b} which is  a complete cosmology that shows excellent agreement with all major cosmological observations.

The first strong evidence for time dilation in type Ia supernovae  was provided by \citet{Leibundgut96} with one supernova and \citet{Goldhaber96} with seven SNe. This was quickly followed by multiple SNe results from \citet{Goldhaber97,Perlmutter99,Goldhaber01}.  These papers record developments in both SNe observations and analysis, the results of which are asserted to provide strong evidence for  an expansion model chiefly because they show that the width of type Ia supernova light curves appears to increase with redshift in good agreement with an expanding model.

The results of this paper are based on the extensive analysis\hspace{0.3ex} of\hspace{0.3ex} type\hspace{0.3ex} Ia\hspace{0.3ex} supernova\hspace{0.3ex} observations\hspace{0.3ex} provided\hspace{0.3ex} by

\hspace{-3.6ex}\citep{Betoule14} (hereafter B14). The raw observations are those used by B14 but without SALT2 calibration. The results of this paper are based on the calibration templates for the SALT2 analysis and the raw type IA supernovae observations described in Section~\ref{s2}.

There is an intrinsic variation of the shape of  light curves of SNe with emitted wavelength, which confounds any redshift variation and needs to be removed in order to measure the  peak luminosity and width for each supernova.  This removal is done  by comparing  the observations of each supernova to  a reference light curve obtained from analyzing all the light curves in the rest-frame (the emitted frame). The usual assumption is made that all the type Ia supernovae are identical irrespective of redshift. The great benefit of this approach is that it provides an estimate of the intrinsic variations in the light curves  that can be subtracted from any observed light curve.

However it is shown  that in general this calibration process  removes all the effects of time dilation and other systematic effects so that the calibrated light curves cannot be used to test cosmologies.  The standard set (SALT2) of templates that provide the reference light curve as a function of rest-frame wave length show direct evidence of  the variations of the original light curve widths with redshift. It shows that the standard deletion of the expected time dilations is not needed and  that the original observations are consistent with a static universe.

The major evidence that supports the hypothesis of a static universe is that a direct  analysis of the raw SNe observations (i.e. without SALT2 calibrations) is that the widths of the SNe light curves do not show time dilation even though there are observed spectral line redshifts.  Furthermore using a static cosmology the absolute magnitudes are independent of redshift. The implication is that the universe is static.

There are two further findings from SNe observations that appear to support the expansion model. First is the apparent dependence of photometric-redshift observations on redshift. These are observations that photometric properties of type Ia supernova spectra, as distinct from spectral wavelength measurements used to determine redshift,  show a redshift dependence. However what they show is a light-curve width dependence not a redshift dependence. Second the age of a spectrum is the number of days between the observation of the spectrum and the epoch of the peak magnitude of the supernova. The ability to determine the age from subtle changes in the spectrum   provides an independent method of estimating the  light-curve width of the supernova. Provided it is not interpreted as a redshift dependence this light-curve width  dependence is consistent with a static cosmology.

\section{Methods}
\label{s2}
\subsection{SALT2 templates}
\label{s2.1}
The B14 calibration method \citep{guy10,Guy07} uses the SALT2 templates (Spectral Adaptive Light-curve Templates) which provide the expected flux density of the supernova light curve as a function of the rest-frame wavelength and the  difference between the observed epoch and the epoch of maximum response. The standard SALT2 template file, $SALT2\_template\_0.dat$, provides the template light curve  for approximately 20 days prior to the maximum and 50 days after the maximum for rest-frame wavelengths from 200 nm to 920 nm in steps of 0.5 nm.  This template file for the JLA (Joint Light-curve Analysis) analysis was taken from the SNANA \citep{Kessler09} website in the directory models$/SALT2/\-SALT2.JLA-B14$.

\subsection{Raw type Ia supernovae light curves}
\label{s2.2}
Recently B14  have provided an update of the \citet{Conley11} analysis with better optical calibrations and more SNe. This JLA (Joint Light-curve  Analysis) list sample has 720 SNe  from the Supernova Legacy Survey (SNLS), nearby SNe (LowZ), the Sloan Digital Sky Survey (SDSS) \citep{Holtzman08,Kessler09} and those revealed by the Hubble Sky Telescope (HST) \citep{Riess07}. The B14 data file provided the supernova name, the redshift, the apparent magnitude and its uncertainty, the stretch parameter ($x_1$)  and its uncertainty, the color parameter (C) and its uncertainty, the host stellar mass and finally the survey number. The major use of the B14 data in this paper is to identify an excellent set of type Ia SNe.

All of the original SNe observations  were retrieved from the SNANA \citep{Kessler09} website using the index  files shown in Table~\ref{tf}. The final column shows the number of B14 SNe that were recovered from each set of files.
\begin{table}[!h]\centering
\caption{Index source files for B14 data}
\label{tf}
\begin{tabularx}{0.485\textwidth}{XP{0.7cm}}
   file                               & count \\ \hline
 {\it \small lcmerge/LOWZ\_JRK07}                         &  48\\
 {\it \small lcmerge/JLA2014\_CSP.LIST}                   &  13\\
 {\it \small lcmerge/JLA2014\_CfAIII\_KEPLERCAM.LIST}     &  32\\
 {\it \small lcmerge/JLA2014\_CfAIII\_4SHOOTER.LIST}      &  21\\
 {\it \small lcmerge/SNLS3year\_JRK07.LIST}               &   2\\
 {\it \small lcmerge/SDSS\_allCandidates+BOSS\_HEAD.FITS} & 369\\
 {\it \small lcmerge/JLA2014\_SNLS.LIST}                  & 238\\
 {\it \small lcmerge/JLA2024\_HST.LIST}                   &   9\\
 {\it \small lcmerge/SDSS\_HOLTZ08}                       &   1\\
\end{tabularx}
\end{table}

For each supernova the following data was extracted: the supernova name and redshift and then for each epoch and for each filter the flux density and its uncertainty. Most of the SNe were observed in four or five filters.  Following B14 the data for the filters $u$ and $U$ was not used. A basic requirement was that there were at least four good epochs that lay between -15 days and +50 days from the peak epoch and that there was at least one observation three days prior to the peak epoch. In addition any flux density that had an uncertainty greater than 30\% of its value was discarded.

The raw SNe observations were analyzed  without using  the SALT2 method.   For each SNe  the fitted parameters were the peak flux density, the epoch of the peak flux density and the relative width of the light curve.  This fitting was done using the  reference light curve provided by \citet{Goldhaber01}.  Then the two scale parameters, the peak flux density and the width, (and the epoch of peak flux density) were determined for each filter. Since these scale parameters are orthogonal they can be determined by separate least squares analysis. For the peak flux density this done by  minimizing the average square of the difference between the observed flux density at each epoch and the height of the reference curve at that epoch multiplied by the peak flux density and divided by the uncertainty in the flux density. In other words a minimum $\chi^2$ analysis.  A similar process was used for the widths and the  peak flux density epoch with the expected epoch being the epoch of the reference curve times the peak flux density that was the same as the observed flux density. The central epoch was common to all filters whereas there was a peak flux density and width provided for each filter. Although the flux density measurement and the width measurement are almost orthogonal there was a small interaction which was eliminated by repeating the  fitting of these parameters for each supernova until they were unchanging.

For each supernova peak flux density estimation all the individual contributions to the $\chi^2$ sum  were put in an array. Any value whose absolute difference from the mean of the other values was greater than five times the rms of the other values was rejected. The values were rejected in turn, starting with the largest discrepancy and finishing when the were no more rejections or there were only four values left. The overall result was that  5.6\% of the flux density measurements that were rejected.  Note that if all the outliers were included there were only small changes to the final results and no changes to any conclusions.

Since the use of uncertainties made negligible difference to the final results and for simplicity all regressions using these peak flux densities and widths were done without weight factors.

\section{Results}
\label{3a}
\subsection{A problem with rest-frame calibration}
\label{s3}
Assuming that the intrinsic shape of the light curve for supernova type Ia is the same at all redshifts then the only effects produced by  cosmology are variations in the scaling constants, the peak flux density and the width.  All information about the cosmology is contained in the  variation of these scaling parameters of the light curves of supernovae at different redshifts.

Since the generation of the type Ia supernova light curve is poorly understood the main method of calibration of a test observation is to compare the test light curve with the average of known light curves. This is necessary because there are intrinsic, wavelength dependent variations in the light curves.  Ideally this would be done with many light curves observed near the same redshift.  However there are not enough reference light curves for this to done for a wide range of redshifts. A second problem with this ideal calibration method is that a test light curve calibrated with the local reference light curves would not contain any cosmological information. For example both  test light curve and the ideal calibration light curve would have the same expected width and therefore the calibrated test light curve would have a standard width independent of redshift and subject only to measurement uncertainties. The SALT2 calibration method and equivalent rest-frame calibration methods are brilliant attempts to solve both of these problems. Unfortunately they solve the problem of having a common reference light curve for all redshifts that cancels the intrinsic effects but at the cost of removing  all the cosmological information.

For simplicity let us assume that the characteristic of interest is the width of the light curve. If we assume that the intrinsic light curve of a type Ia supernova is the same at all redshifts then the effect of any cosmological model is to determine scaling of the light curve as a function of redshift. In particular this scaling will change the observed width  and height (peak flux density) of the light curve. If $W(\alpha)$ is the intrinsic width of the light curve at the rest-frame wavelength, $\alpha$, and let  the expected width at an observed  wavelength $\beta=(1+z)\alpha$  be $V(z)$.  Note that because  the supernova are identical at all redshifts then all observations of a particular supernova must have the same width independent of $\beta$ then $V(z)$  is only a function of $z$.

Classically this is just like the Doppler effect in that this redshift is applicable to every time interval. Thus if the universe is expanding then $V(z)/W(\alpha))=(1+z)$. However quantum mechanics shows that the spectral line wavelengths are a function of photon energy and therefore the spectroscopic redshift is not necessarily the same as the time dilation for macroscopic time intervals. Thus it could be produced by a "tired light" process and not by universal expansion. However, by definition, the redshift dependence of time dilation can only be produced by universal expansion.

The expected  width, $V(z)$, observed by a telescope with a filter gain function $g(\beta)$ at redshift $z$ for a particular supernova is the integral over this gain function, namely
\begin{equation}
\label{e2}
V(z)= \int{W(\alpha)g(\beta)d\beta}.
\end{equation}
Then if $V^*(z)$ is the observed width the calibrated width is
\begin{equation}
\label{f3}
V_{calibrated}= V^*(z)/V(z).
\end{equation}

Now suppose that intrinsic width is a power  law function of the wavelength with an exponent $\gamma$ so that
\begin{equation}
\label{e3}
W(\alpha)=A\alpha^\gamma,
\end{equation}
where $A$ is a constant. Since $\alpha=\beta/(1+z)$ equation~\ref{e2} can be written as
\begin{equation}
\label{e4}
V(z)= A(1+z)^{-\gamma}\int{\beta^{\gamma}g(\beta)d\beta} = B(1+z)^{-\gamma},
\end{equation}
where $B$ is defined by
\begin{equation}
\label{e5}
B= A\int{\beta^{\gamma} g(\beta)d\beta}.
\end{equation}
Clearly $B$ is a constant that  depends only on $\gamma$ and the  filter characteristics. Thus if the intrinsic width is a power law of the rest-frame wavelength with exponent $\gamma$ this produces an observed width that is proportional to $(1+z)^{-\gamma}$. For a supernova at redshift $z$  the predicted width is
\begin{equation}
\label{e6}
V(z)= B(1+z)^{-\gamma}.
\end{equation}

Note the reversal of sign of the exponent  $\gamma$ between wavelength dependence and redshift dependence. Since the integration is linear and since any intrinsic distribution, $W(\alpha)$, can be expressed as sum of powers of $\alpha$ then it follows that $V(z)$ will be a sum of power laws  which means that any intrinsic  rest-frame wavelength distribution can be fully represented by an observed  redshift distribution. Thus there is a direct correspondence between the power laws as a function of wavelength in the rest-frame and the power laws as a function of redshift in the observations. This is similar to a Fourier transform which provides two ways of interpreting the same data.

The process of determining $W(\alpha)$ from many observations of $V(z)$ by some de-convolution process is  difficult and not necessarily unique. However a simple  method of doing this is to compute a sum of power laws that accurately represent the redshift distribution and then use the equations above to produce the intrinsic wavelength distribution in the rest-frame.

It is clear that having computed $W(\alpha)$ the reconstruction of say, $V(z)$, from $W(\alpha)$ must be identical to the original $V(z)$.  Thus apart from measurement uncertainties the reference width for any redshift must equal the expected  width and thus the  calibrated width  will be (within measurement uncertainties) a constant. Thus there will not be any cosmological information in this calibrated width. In effect the SALT2 (and similar) calibration methods cannot distinguish between intrinsic and redshift variations of width. It removes both of them. Clearly this conclusion is applicable to absolute magnitudes.

To summarize an  important consequence of this analysis is that any rest-frame calibration method will transmit all of any systematic variation in width or peak flux density to the rest-frame templates. Then use of these templates to calibrate  the observations of a test supernova will mean that the systematic variation will cancel its occurrence in the observations. Thus any cosmological information will be eliminated from the calibrated width and peak flux density. Consequently  any rest-frame  calibration method like SALT2 will achieve its aim of eliminating all the intrinsic variations but at the cost of elimination of all of the cosmological information.

A particular example is the standard calibration using the SALT2 method. In this case whether or not there is time dilation or whether or not the epoch differences are divided by $(1+z)$ the calibrated widths will not show any variation of widths with redshift. In practice the estimation of auxiliary parameters may cause some small variations. Thus the fact that the stretch parameters show little variation with redshift is not surprising. Also the multiplication of the stretch factors by $(1+z)$  to get the "observed width" is unwarranted.

\subsection{SALT2 templates}
\label{s3.1}
\begin{figure}[!t]
\includegraphics[width=\columnwidth]{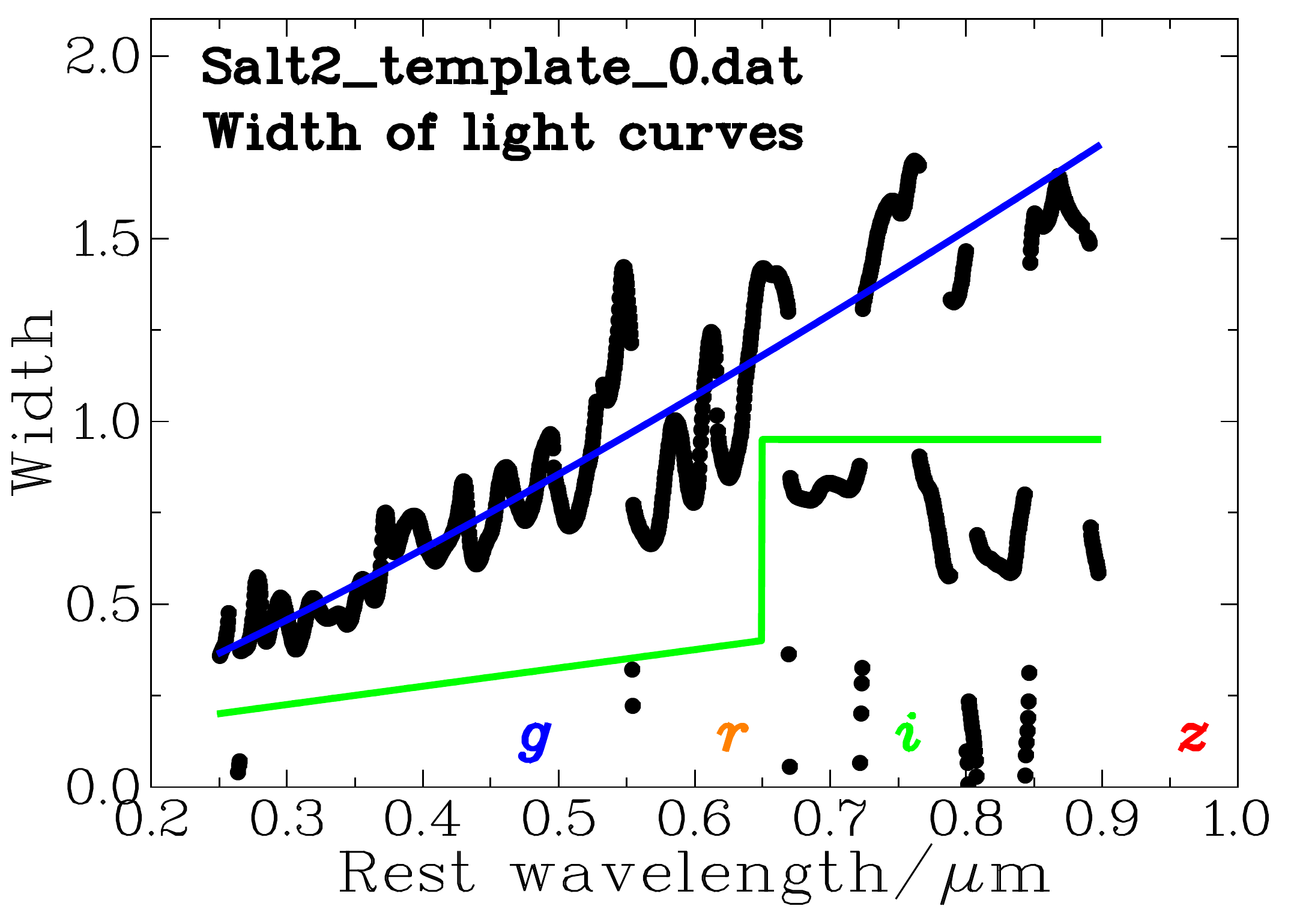}
\caption{Plot of the relative  width  of  the SALT2 template SNe light curves as a function of rest-frame wavelength $\lambda$ shown as solid black circles.  The blue line shows  the fitted  wavelength dependence equal to $1.927\lambda^{1.199}$ with $\lambda$ in nm. As explained in the text only points above the green line  have been used.  The average rest-frame wavelength  response for each filter is shown at the bottom of the figure  for the $griz$ filters. }
\label{fig1}
\end{figure}

The width of each SALT2 template light curve was taken to be  the distance between the two half peak values divided by 22.4 days. In Figure~\ref{fig1} the black points show the width of the light curves in the SALT2 templates used by B14 as a function of rest-frame wavelength. The average wavelength of the $griz$ (These are the names for standard the telescope wavelength filters.) filters are shown at the bottom of the figure. Note that in places the width is poorly determined. Since the longer wavelength results are mainly determined by the nearby SNe these poor regions roughly correspond to the wavelength regions between the filters. The green line in Figure~\ref{fig1} shows the upper limit of where the fitted widths are valid. After all if this variation in widths is of cosmological origin it must be a smooth function of wavelength. At shorter wavelengths there is a smoothing effect due to the spread of redshifts.

A power law fit for widths that were above the green line shown in Figure~\ref{fig1} was done to get the regression equation
\begin{equation}
W(\alpha)= (1.927\pm 0.010)\alpha^{(1.199\pm0.014)},
\end{equation}
which is shown as a solid blue line in Figure~\ref{fig1}  and it shows  a  good fit to the data.

As a check on this analysis the expected observed width $V(z)$ was computed from a rest-frame spectrum with an exponent of $\gamma=1.199$ using $griz$ filter gain curves and equation~\ref{e2}. The computed widths were proportional to $(1+z)^{-1.227\pm 0.027}$ which is a check on equation~\ref{e4} and is in good agreement with the original observed value.

The first step in the SALT2 calibration process is to remove the assumed time dilation by dividing all epoch differences by $(1+z)$. If the universe is static there is no time dilation and the input observations to the rest of the SALT2 analysis would give a width that varies as $(1+z)^{-1}$. Section~\ref{s3} shows that this would produce an intrinsic wavelength dependence proportional to $\lambda$. Hence an  exponent of 1.199 implies that intrinsic power law dependence would  be proportional to $\lambda^{ 0.199}$  or equivalently  the raw light curves would have widths that  would be proportional to $(1+z)^{-0.199}$.
This shows that with this small intrinsics wavelength dependence  of the widths this template results could be consistent with a static universe.

If the standard time dilation is present then the rest-frame spectrum must have a dependence of  $\lambda^{1.199}$ which seems rather large and should be obvious in the widths between different filters for low redshift SNe. This point is discussed further in Section~\ref{s4.1}.

\subsection{The lack of time dilation in type Ia supernovae}
\label{s4.1}
This section is based on the analysis of the raw supernova observations described in section~\ref{s2.2}.
In order to investigate whether  there is a systematic variation of width with the wavelength  a regression was done for the widths for each filter as a function of redshift.   None of the filters showed a significant variation of width with redshift.  The asymptotic  width,
$V_0$ (the value of the regression equation at $z=0$), is shown in column 5 of  Table~\ref{t0}. The last column in Table~\ref{t0} shows the absolute magnitude at zero redshift, $M_0$, for each filter for the static model (section~\ref{s4.2}).
\begin{table}[!h]\centering
\caption{Asymptotic widths and peak absolute magnitudes}
\label{t0}
\begin{tabularx}{0.485\textwidth}{YP{1.7cm}YP{2cm}Y}
 filter &  Mean $\beta$ & No. & $V_0$ & $M_0$  \\ \hline
$B$  & 0.436 $\mu$m &  22 & $(0.784 \pm 0.024)$ &-18.519\\
$V$  & 0.541 $\mu$m &  22 & $(1.014 \pm 0.022)$ &-18.928\\
$R$  & 0.619 $\mu$m &  34 & $(1.051 \pm 0.021)$ &-19.030\\
$I$  & 0.750 $\mu$m &  32 & $(1.107 \pm 0.026)$ &-18.813\\
$g$  & 0.472 $\mu$m & 439 & $(0.873 \pm 0.007)$ &-19.088\\
$r$  & 0.619 $\mu$m & 530 & $(1.041 \pm 0.006)$ &-19.098\\
$i$  & 0.750 $\mu$m & 531 & $(1.051 \pm 0.007)$ &-18.788\\
$z$  & 0.958 $\mu$m & 300 & $(1.013 \pm 0.015)$ &-18.484\\
\end{tabularx}
\end{table}

Although there is no significant dependence on redshift there is a  dependence of the asymptotic  value on the  wavelength. Therefore in order to put the width measurements onto a common reference frame all the widths were divided by the asymptotic width for the same filter (column 4). This will also remove some of the effects of the supernovae light curves having an intrinsic width  as a function of wavelength.

In Section~\ref{s3.1} it shown that if there was time dilation then the width spectrum of the SALT2 templates must have an intrinsic spectrum proportional to $\lambda^{1.199}$. Table~\ref{t0} shows that it is the $B$ and $g$ filters that are most discrepant. Using the  asymptotic widths and average wavelengths from Table~\ref{t0} for filters $griz$  the fitted power law has an exponent of $0.33\pm0.12$ which is significantly less than the exponent of 1.199 required by the standard expansion model but is consistent with the exponent of 0.199 required by the static model. This shows support for a static universe.

\begin{figure}
\includegraphics[width=\columnwidth]{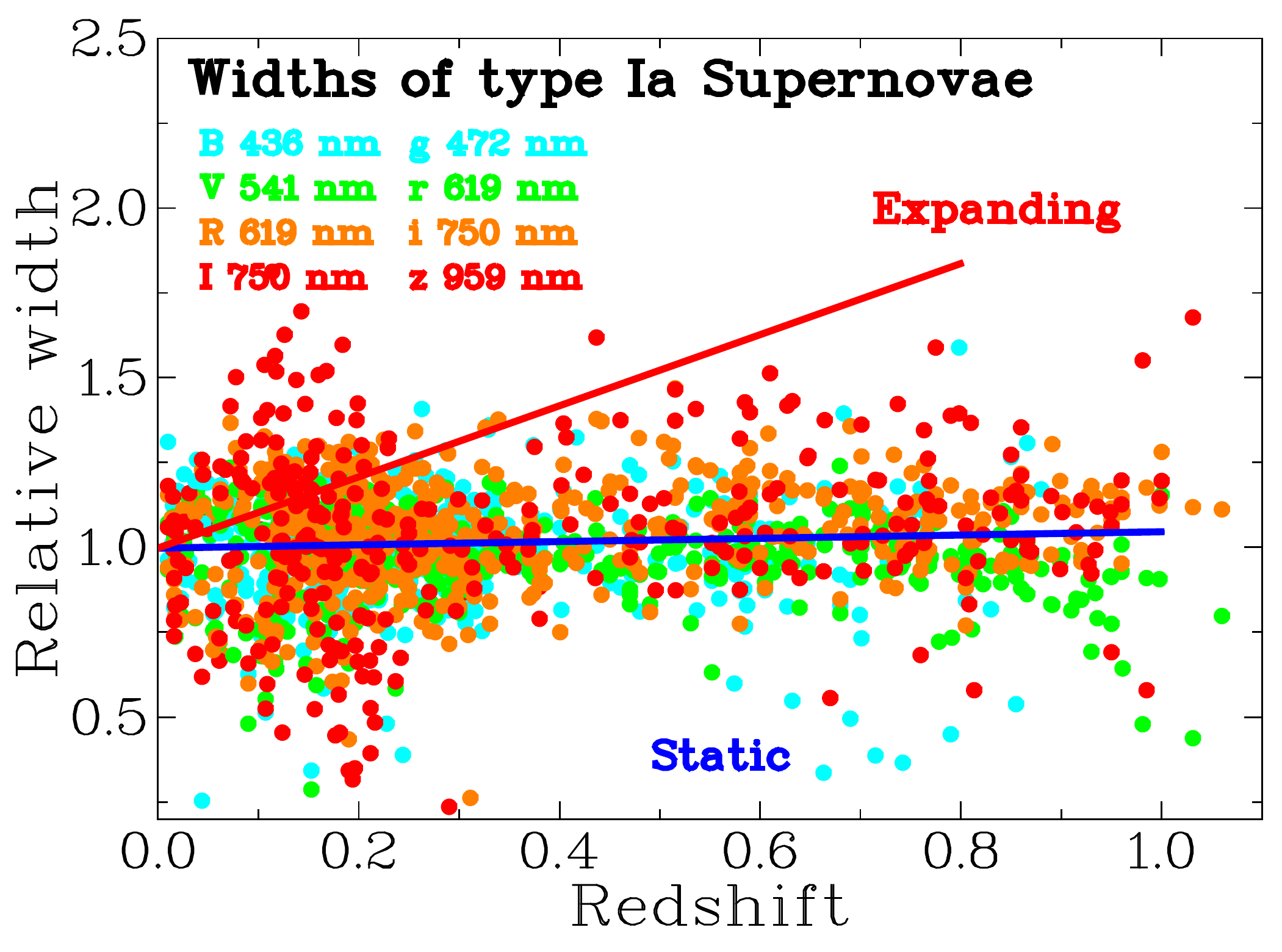}
\caption{\label{fig2} Plot of the measured width for the light curve for each supernova and filter a function of redshift. The blue line shows the regression line given in equation~\ref{e3.1} and the red line shows the expected width ($w=1+z$) if there was time dilation. The legend shows the plotted color for each filter.}
\end{figure}

Figure~\ref{fig2} shows a plot of all the widths as a function of redshift.  The assumption made here is that the asymptotic corrections will remove most of the intrinsic variations. Any remaining intrinsic variation will just increase the scatter of the redshifts and possibly produce an extra redshift dependence.  The  regression as a function of redshift for the 1910 values is
\begin{equation}
\label{e3.1}
V(z)= (0.998\pm0.004) + (0.048\pm0.016)z.
\end{equation}
Although the slope is just at the 3$\sigma$ level it should be noted that the observed intrinsic width discussed in section~\ref{s3.1} corresponds to a slope of 0.199 and this intrinsic width is only partially removed by dividing the observed widths by the asymptotic width. Thus the slope in equation~\ref{e3.1} is consistent with zero and completely inconsistent with the expanding slope of one (63$\sigma$).

Clearly the regression equation and the distribution of points in Figure~\ref{fig2} show that the widths of the light curves for these supernovae are consistent with no time dilation. If there is time dilation then there the intrinsic width spectrum $V(z)$ must have a power law spectrum with an exponent very close to one. The type Ia supernovae light curve is produced from a rapidly expanding gas cloud and is complicated function of area, chemical composition, temperature and transparency of the expanding gas cloud and it is difficult to see why the wave length dependence should have an exponent so close to one. Furthermore the evidence from the SALT2 templates and the asymptotic widths show that it has an equivalent dependence of about $(1+z)^-0.2)$ which shows that it is extremely  unlikely that it can explain the lack of time dilation.

Since time dilation is the main defining characteristic of an expanding universe the conclusion is that the universe does not show the standard time dilation  and these results are consistent with a static universe.

\subsection{Type Ia supernovae magnitudes}
\label{s4.2}
The analysis for magnitudes is more complicated than that for widths in that a  distance modulus derived from a cosmological model must be used in order to obtain the absolute magnitudes. For the static cosmological  model the distance modulus (equation~\ref{a6}) for curvature cosmology is used here because it shows excellent results from quasar observations. For the standard expansion cosmology the distance modulus give in Appendix B (equation~\ref{b2}) is used.

For each filter the asymptotic value  of the absolute magnitude at zero red shift (shown in Table~\ref{t0} as $M_0$) minus $M_0$ was used to provide a common reference which was subtracted from each of the observed absolute magnitudes.
For all  for 566 SNe the regression results for the absolute-peak magnitudes as a function of redshift  are shown Table~\ref{t1} for both the static and expanding models.  Note that the equation for the static distance modulus (first given \citet{Crawford95b})  has no free parameters and it has not been adjusted in any way to suit SNe observations.

The interesting result is that the regression for the static model has negligible slope but that for the expanding model has a significantly  larger slope which shows strong support for the static universe and very poor support for an expanding universe. It could be that the intrinsic peak flux density has a dependence that accurately mirrors the standard expansion model distance modulus so that its effects are nullified (see section~\ref{s5}). But  this cannot explain by the static model distance modulus has such an excellent result.
\begin{table}[!h]\centering
\caption{Absolute peak magnitudes verses redshift}
\label{t1}
\begin{tabularx}{0.485\textwidth}{P{1.2cm}YY}
 model        &  regression  equation                        \\ \hline
static        &  $(-18.965 \pm 0.015) - (0.121 \pm 0.058) z$ \\
expanding     &  $(-18.966 \pm 0.015) - (0.585 \pm 0.058) z$ \\
\end{tabularx}
\end{table}
\begin{figure}[!h]\centering\capstart
\includegraphics[width=0.485\textwidth]{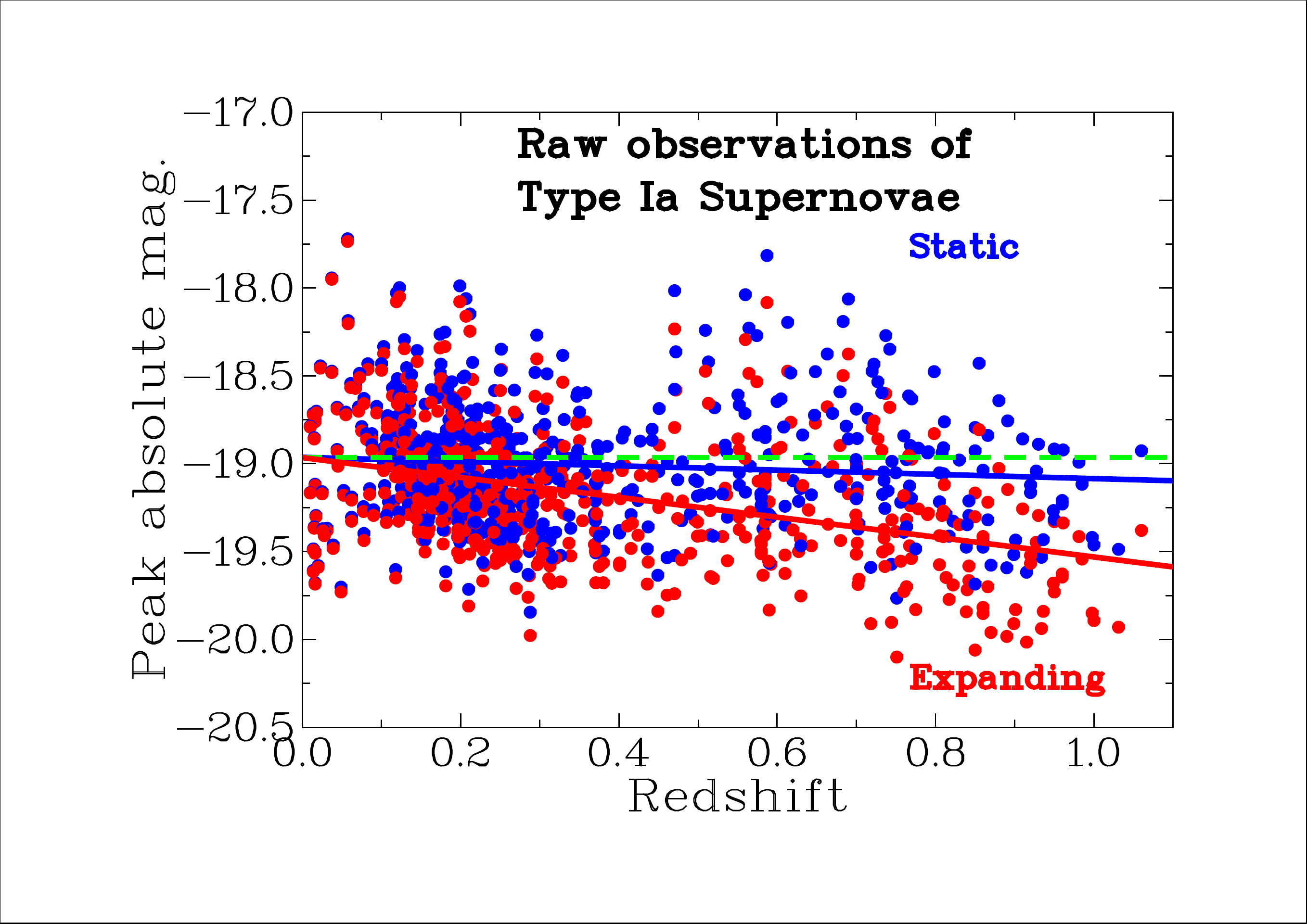}
\caption{Plot of the measured absolute magnitude for each supernova as a function of redshift both static (blue points) and expanding (red points) cosmologies. The blue line shows the regression line given  as the first line in Table~\ref{t1} and the red line shows the expected absolute magnitude if there was an expanding universe (second line in Table~\ref{t1}).}
\label{fig3}
\end{figure}

\subsection{The Phillip's relation}
\label{s5}
\citet{Phillips93} found that the absolute peak-luminosities of SNe appear to be  tightly correlated with the rate of decline of the B light curve. This correlation may be interpreted as  being between peak luminosity and  light curve width is that the more luminous SNe are also wider.  The Phillips relation is intrinsic to the SNe and thus independent of any cosmological model.

From B14 and \citet{Conley11}  we get $\Delta m\approx-1.44\Delta V$.  Since the static distance modulus is a good model for the peak absolute magnitudes we can use it as a reference in order to compute $\Delta m$ for the expanding model. A regression fit for the difference between the two distance moduli shows that to high accuracy  $\Delta m = 0.013 - 0.448z$.  In this context $\Delta V=z$ which results in  $\Delta m \approx -0.448\Delta V$. What is interesting is that apart from the different coefficients this is the same as for the B14 result.

Although the coefficients are different this common dependence suggests that the Phillip's relation may be a consequence of trying to reconcile the anomalous widths and magnitudes within the paradigm of an expanding universe.

\subsection{Apparent extra wide SNe}
\label{s6}
There are reports of SNe with very wide light curves that are inconsistent with a static model. An example of apparent extra wide light curves is the 14 light curves shown by \citet{Suzuki12}. For these supernovae the redshifts vary from 0.623 to 1.415 and the light curve (expansion model) widths vary from 1.48 to 2.43.  Although the  stretch parameter widths vary from 0.77 to 1.09 and are well within the expected range, section~\ref{s3} shows that these stretch factors are invalid and the multiplication of the stretch factors by $(1+z)$ to get the widths is unwarranted.

\subsection{Photometric redshift}
\label{s7}
The redshift of a supernovae can be estimated by comparing its wavelength spectrum  to the average  rest-frame wavelength spectrum as distinct from spectral wavelength measurements used to determine redshift.  For example ever since \citet{Tripp98} showed that there was a correlation between redshifts of SNe and their color index B-V there has been a considerable effort \citep{Howell07, Bazin11, Guy07, Mohlabeng13, Wang13} to use this correlation in order to develop a predictor of the redshift from photometric measurements. However in a static universe although this is a valid estimate of the light-curve widths it is not evidence for time dilation. It is only in the expansion model that the  time dilation  is related to redshift.

\subsection{Spectroscopic ages}
\label{s8}
Another example estimating the width of the supernovae light curve is from spectroscopic ages.  SNe show a consistent variation in characteristics of their spectra with the number of days before and after the maximum. This variation is due to changes in composition, changes in the velocity of the ejecta and the depth of penetration of the ejecta. \citet{Blondin08} have made a comprehensive analysis of these spectra for both local SNe and 13 high redshift SNe that shows that the age (the position in the light curve from the position of the peak luminosity) of a spectrum can be estimated to within 1-3 days. If there are two or more spectra the aging rate and therefore the width can be estimated. In their analysis they explicitly assumed that this width dependence was a measure of redshift which is true  only for the expansion model.

\section{Discussion and conclusions}
\label{s9}
It is  shown that in general  the analysis of SNe calibrated by SALT2 or a similar rest-frame method cannot be used to obtain cosmological information.  Thus the characteristics of calibrated light curves obtained using SALT2 (or equivalent) method are independent of what cosmology is present or what cosmology is assumed.  Examination of the reference light curve widths  in the B14 SALT2 templates shows that are what would consistent be expected if the universe was static. This is strong evidence for a static cosmology.

The normal Hubble variation of observed wavelength with redshift is well established and in an expanding universe time dilation should show the identical dependence. Observations of  type Ia supernovae are  one of the few observations that can directly show time dilation. However this paper shows that raw  observations of the light curve widths of type Ia supernovae do not show the effects of time dilation. In addition using a static cosmological model it was shown that the peak absolute magnitude if these raw observations is independent of redshift. Both of these results show strong support for a static universe.

In \citep{Crawford09b} I have argued that Curvature Cosmology has a much better agreement with cosmological observations than the standard expansion model. This paper completes this work by showing that this is also true for type Ia supernovae light curves. One interesting result from this earlier work is that it can explain the velocity dispersion of galaxies in the Coma cluster without requiring dark matter.

A definitive test of a static universe would be to completely repeat the SALT2 calibration procedure using a static model. The test will be that the rest-frame calibration light curves will show only  variations with redshift that are due to intrinsic wavelength effects.

The most important conclusions for this paper are:
\begin{itemize}
\item[\em $\bullet$\/]	 Type Ia supernovae light curves calibrated with SALT2 or other rest-frame methods cannot provide tests of cosmology.
\item[\em $\bullet$\/]	 Widths of  SALT2 template light curves are consistent with a static universe.
\item[\em $\bullet$\/]	 Type Ia supernovae do not show time dilation.
\item[\em $\bullet$\/]	 The peak magnitudes of  type Ia supernovae are consistent with a static universe.
\end{itemize}

\appendix
\section{Static Cosmology}

The static cosmology used here is Curvature Cosmology \citep{Crawford09b} that is a complete cosmology that shows excellent agreement with all major cosmological observations. In particular it can explain the exceptionally large velocity dispersion of galaxies in a cluster without requiring dark matter. The distance modulus has excellent agreement with quasar observations. The geometry is that of a three  dimensional surface of a four dimensional hyper sphere. For this geometry the area of a three dimensional sphere with radius  $r=R\chi$ \citep{Crawford09b} where $\chi =\ln(1+z)/\sqrt{3}$ and the surface area is
\begin{equation}
A(r) = 4\pi R^2 \sin ^2 (\chi ).
\end{equation}
From this we get the equation for the distance-modulus, ($\mu=m-M$),  which is
\begin{equation}
\label{a6}
\mu_{static}= 5\log \left[{\frac{\sqrt{3} \sin(\chi)}{h}} \right] + 2.5\log(1+z) + 42.384\\
\end{equation}
where $h=H/100 {\,kms}^{-1}{\,Mpc}^{-1}$.

\section{Expansion model functions}
The equations needed for the modified $\Lambda$-CDM model \citep{Hogg99,Goliath01,Barboza08}, with $\Omega_M=0.27$,
$\Omega_K=0$  and where $h$ is the reduced Hubble constant, are listed below. The symbol $w^*$ is used for the acceleration parameter in order to avoid confusion with the width, $w$. These equations depend on the function $E(z)$ defined here by
\begin{equation}
\label{b1}
E(z) = \int_0^z \frac{dz}{\sqrt{\Omega_M (1+z)^3+(1-\Omega_M)(1+z)^{(1+w^*)}}}.
\end{equation}
The distance modulus is
\begin{equation}
\label{b2}
\mu_{exp}(z)=5\log(E(z)(1+z)/h)+ 42.384.
\end{equation}
The equation of state parameter $w^*$ in the expansion model distance modulus is included to investigate the effects of including the cosmological constant.  \citet{Conley11} found that  the  parameter, $w^*$,  has a value $w^*=-0.91$, whereas \citet{Sullivan11} found that $w^*=-1.069$.   Although its actual value is not critical for this paper the value of $w^*$ is chosen to be  $w^*=-1.11$.


\begin{thebibliography}{50}
\expandafter\ifx\csname natexlab\endcsname\relax\def\natexlab#1{#1}\fi
\providecommand{\url}[1]{\href{#1}{#1}}

\bibitem[{{Barboza} \& {Alcaniz}(2008)}]{Barboza08}
{Barboza}, E.~M., \& {Alcaniz}, J.~S. 2008, Physics Letters B, 666, 415--419.

\bibitem[{{Bazin} {et~al.}(2011){Bazin}, {Ruhlmann-Kleider},
  {Palanque-Delabrouille}, {Rich}, {Aubourg}, {Astier}, {Balland}, {Basa},
  {Carlberg}, {Conley}, {Fouchez}, {Guy}, {Hardin}, {Hook}, {Howell}, {Pain},
  {Perrett}, {Pritchet}, {Regnault}, {Sullivan}, {Fourmanoit},
  {Gonz{\'a}lez-Gait{\'a}n}, {Lidman}, {Perlmutter}, {Ripoche}, \&
  {Walker}}]{Bazin11}
{Bazin}, G., {Ruhlmann-Kleider}, V., {Palanque-Delabrouille}, N., {et~al.}
  2011, \aap, 534, A43.

\bibitem[{{Betoule} {et~al.}(2014){Betoule}, {Kessler}, {Guy}, {Mosher},
  {Hardin}, {Biswas}, {Astier}, {El-Hage}, {Konig}, {Kuhlmann}, {Marriner},
  {Pain}, {Regnault}, {Balland}, {Bassett}, {Brown}, {Campbell}, {Carlberg},
  {Cellier-Holzem}, {Cinabro}, {Conley}, {D'Andrea}, {DePoy}, {Doi}, {Ellis},
  {Fabbro}, {Filippenko}, {Foley}, {Frieman}, {Fouchez}, {Galbany}, {Goobar},
  {Gupta}, {Hill}, {Hlozek}, {Hogan}, {Hook}, {Howell}, {Jha}, {Le Guillou},
  {Leloudas}, {Lidman}, {Marshall}, {M{\"o}ller}, {Mour{\~a}o}, {Neveu},
  {Nichol}, {Olmstead}, {Palanque-Delabrouille}, {Perlmutter}, {Prieto},
  {Pritchet}, {Richmond}, {Riess}, {Ruhlmann-Kleider}, {Sako}, {Schahmaneche},
  {Schneider}, {Smith}, {Sollerman}, {Sullivan}, {Walton}, \&
  {Wheeler}}]{Betoule14}
{Betoule}, M., {Kessler}, R., {Guy}, J., {et~al.} 2014, \aap, 568, A22.

\bibitem[{{Blondin} {et~al.}(2008){Blondin}, {Davis}, {Krisciunas}, {Schmidt},
  {Sollerman}, {Wood-Vasey}, {Becker}, {Challis}, {Clocchiatti}, {Damke},
  {Filippenko}, {Foley}, {Garnavich}, {Jha}, {Kirshner}, {Leibundgut}, {Li},
  {Matheson}, {Miknaitis}, {Narayan}, {Pignata}, {Rest}, {Riess}, {Silverman},
  {Smith}, {Spyromilio}, {Stritzinger}, {Stubbs}, {Suntzeff}, {Tonry},
  {Tucker}, \& {Zenteno}}]{Blondin08}
{Blondin}, S., {Davis}, T.~M., {Krisciunas}, K., {et~al.} 2008, \apj, 682, 724--736.

\bibitem[{{Conley} {et~al.}(2011){Conley}, {Guy}, {Sullivan}, {Regnault},
  {Astier}, {Balland}, {Basa}, {Carlberg}, {Fouchez}, {Hardin}, {Hook},
  {Howell}, {Pain}, {Palanque-Delabrouille}, {Perrett}, {Pritchet}, {Rich},
  {Ruhlmann-Kleider}, {Balam}, {Baumont}, {Ellis}, {Fabbro}, {Fakhouri},
  {Fourmanoit}, {Gonz{\'a}lez-Gait{\'a}n}, {Graham}, {Hudson}, {Hsiao},
  {Kronborg}, {Lidman}, {Mourao}, {Neill}, {Perlmutter}, {Ripoche}, {Suzuki},
  \& {Walker}}]{Conley11}
{Conley}, A., {Guy}, J., {Sullivan}, M.,  {Regnault}, N., {Astier}, P.,{ Balland}, C. {et~al.} 2011, \apjs, 192(1), 1.

\bibitem[{{Crawford}(1995)}]{Crawford95b}
{Crawford}, D.~F. 1995, \apj, 441(2), 488--493.

\bibitem[{{Crawford}(2009)}]{Crawford09b}
{Crawford}, D.~F. 2009, arXiv:\url{http://arxiv.org/abs/1009.0953}

\bibitem[{{Goldhaber}(1997)}]{Goldhaber97}
{Goldhaber}, G. 1997, in NATO ASIC Proc. 486: Thermonuclear Supernovae, ed.
  P.~{Ruiz-Lapuente}, R.~{Canal}, \& J.~{Isern}, 777

\bibitem[{{Goldhaber} {et~al.}(1996){Goldhaber}, {Boyle}, {Bunclark}, {Carter},
  {Couch}, {Deustua}, {Dopita}, {Ellis}, {Filippenko}, {Gabi}, {Glazebrook},
  {Goobar}, {Groom}, {Hook}, {Irwin}, {Kim}, {Kim}, {Lee}, {Matheson},
  {McMahon}, {Newberg}, {Pain}, {Pennypacker}, {Perlmutter}, \&
  {Small}}]{Goldhaber96}
{Goldhaber}, G., {Boyle}, B., {Bunclark}, P., {et~al.} 1996, Nuclear Physics B
  Proceedings Supplements, 51, 123--127.

\bibitem[{{Goldhaber} {et~al.}(2001){Goldhaber}, {Groom}, {Kim}, {Aldering},
  {Astier}, {Conley}, {Deustua}, {Ellis}, {Fabbro}, {Fruchter}, {Goobar},
  {Hook}, {Irwin}, {Kim}, {Knop}, {Lidman}, {McMahon}, {Nugent}, {Pain},
  {Panagia}, {Pennypacker}, {Perlmutter}, {Ruiz-Lapuente}, {Schaefer},
  {Walton}, \& {York}}]{Goldhaber01}
{Goldhaber}, G., {Groom}, D.~E., {Kim}, A.,  {Aldering}, G., {Astier}, P., {Conley}, A. {et~al.} 2001, \apj, 558(1), 359--368.

\bibitem[{{Goliath} {et~al.}(2001){Goliath}, {Amanullah}, {Astier}, {Goobar},
  \& {Pain}}]{Goliath01}
{Goliath}, M., {Amanullah}, R., {Astier}, P., {Goobar}, A., \& {Pain}, R. 2001,
  \aap, 380, 6--18.

\bibitem[{{Guy} {et~al.}(2007){Guy}, {Astier}, {Baumont}, {Hardin}, {Pain},
  {Regnault}, {Basa}, {Carlberg}, {Conley}, {Fabbro}, {Fouchez}, {Hook},
  {Howell}, {Perrett}, {Pritchet}, {Rich}, {Sullivan}, {Antilogus}, {Aubourg},
  {Bazin}, {Bronder}, {Filiol}, {Palanque-Delabrouille}, {Ripoche}, \&
  {Ruhlmann-Kleider}}]{Guy07}
{Guy}, J., {Astier}, P., {Baumont}, S., { Hardin}, D., {Pain}, R., {Regnault}, N. {et~al.} 2007, \aap, 466, 11--21.

\bibitem[{{Guy} {et~al.}(2010){Guy}, {Sullivan}, {Conley}, {Regnault},
  {Astier}, {Balland}, {Basa}, {Carlberg}, {Fouchez}, {Hardin}, {Hook},
  {Howell}, {Pain}, {Palanque-Delabrouille}, {Perrett}, {Pritchet}, {Rich},
  {Ruhlmann-Kleider}, {Balam}, {Baumont}, {Ellis}, {Fabbro}, {Fakhouri},
  {Fourmanoit}, {Gonz{\'a}lez-Gait{\'a}n}, {Graham}, {Hsiao}, {Kronborg},
  {Lidman}, {Mourao}, {Perlmutter}, {Ripoche}, {Suzuki}, \& {Walker}}]{guy10}
{Guy}, J., {Sullivan}, M., {Conley}, A.,  {Regnault, N.}, {Astier}, P., {Balland}, C {et~al.} 2010, \aap, 523, A7.

\bibitem[{{Hogg}(1999)}]{Hogg99}
{Hogg}, D.~W. 1999, ArXiv Astrophysics e-prints, arXiv:astro-ph/9905116

\bibitem[{{Holtzman} {et~al.}(2008){Holtzman}, {Marriner}, {Kessler}, {Sako},
  {Dilday}, {Frieman}, {Schneider}, {Bassett}, {Becker}, {Cinabro}, {DeJongh},
  {Depoy}, {Doi}, {Garnavich}, {Hogan}, {Jha}, {Konishi}, {Lampeitl},
  {Marshall}, {McGinnis}, {Miknaitis}, {Nichol}, {Prieto}, {Riess}, {Richmond},
  {Romani}, {Smith}, {Takanashi}, {Tokita}, {van der Heyden}, {Yasuda}, \&
  {Zheng}}]{Holtzman08}
{Holtzman}, J.~A., {Marriner}, J., {Kessler}, R., {Sako}, M., {Dilday}, B., {Frieman}, J. A. {et~al.} 2008, \aj, 136(6), 2306--2320.

\bibitem[{{Howell} {et~al.}(2007){Howell}, {Sullivan}, {Conley}, \&
  {Carlberg}}]{Howell07}
{Howell}, D.~A., {Sullivan}, M., {Conley}, A., \& {Carlberg}, R. 2007, \apjl,
  667, L37
\newpage
\bibitem[{{Kessler} {et~al.}(2009){Kessler}, {Becker}, {Cinabro}, {Vanderplas},
  {Frieman}, {Marriner}, {Davis}, {Dilday}, {Holtzman}, {Jha}, {Lampeitl},
  {Sako}, {Smith}, {Zheng}, {Nichol}, {Bassett}, {Bender}, {Depoy}, {Doi},
  {Elson}, {Filippenko}, {Foley}, {Garnavich}, {Hopp}, {Ihara}, {Ketzeback},
  {Kollatschny}, {Konishi}, {Marshall}, {McMillan}, {Miknaitis}, {Morokuma},
  {M{\"o}rtsell}, {Pan}, {Prieto}, {Richmond}, {Riess}, {Romani}, {Schneider},
  {Sollerman}, {Takanashi}, {Tokita}, {van der Heyden}, {Wheeler}, {Yasuda}, \&
  {York}}]{Kessler09}
{Kessler}, R., {Becker}, A.~C., {Cinabro}, D., {Vanderplas}, J., {Frieman}, J. A., {Marriner}, J. {et~al.} 2009, \apjs, 185(1), 32--84.

\bibitem[{{Leibundgut} {et~al.}(1996){Leibundgut}, {Schommer}, {Phillips},
  {Riess}, {Schmidt}, {Spyromilio}, {Walsh}, {Suntzeff}, {Hamuy}, {Maza},
  {Kirshner}, {Challis}, {Garnavich}, {Smith}, {Dressler}, \&
  {Ciardullo}}]{Leibundgut96}
{Leibundgut}, B., {Schommer}, R., {Phillips}, M., {Riess}, A., {Schmidt}, B., {Spyromilio}, J.{et~al.} 1996, \apjl, 466,
  L21.

\bibitem[{{Mohlabeng} \& {Ralston}(2013)}]{Mohlabeng13}
{Mohlabeng}, G.~M., \& {Ralston}, J.~P. 2013, ArXiv e-prints, arXiv:1303.0580

\bibitem[{{Perlmutter} {et~al.}(1999){Perlmutter}, {Aldering}, {Goldhaber},
  {Knop}, {Nugent}, {Castro}, {Deustua}, {Fabbro}, {Goobar}, {Groom}, {Hook},
  {Kim}, {Kim}, {Lee}, {Nunes}, {Pain}, {Pennypacker}, {Quimby}, {Lidman},
  {Ellis}, {Irwin}, {McMahon}, {Ruiz-Lapuente}, {Walton}, {Schaefer}, {Boyle},
  {Filippenko}, {Matheson}, {Fruchter}, {Panagia}, {Newberg}, {Couch}, \&
  {Supernova Cosmology Project}}]{Perlmutter99}
{Perlmutter}, S., {Aldering}, G., {Goldhaber}, G., {Knop}, R. A., {Nugent}, P., {Castro}, P. G {et~al.} 1999, \apj, 517(2),
  565--586.

\bibitem[{{Phillips}(1993)}]{Phillips93}
{Phillips}, M.~M. 1993, \apjl, 413(2), L105-L108.

\bibitem[{{Riess} {et~al.}(2007){Riess}, {Strolger}, {Casertano}, {Ferguson},
  {Mobasher}, {Gold}, {Challis}, {Filippenko}, {Jha}, {Li}, {Tonry}, {Foley},
  {Kirshner}, {Dickinson}, {MacDonald}, {Eisenstein}, {Livio}, {Younger}, {Xu},
  {Dahl{\'e}n}, \& {Stern}}]{Riess07}
{Riess}, A.~G., {Strolger}, L.-G., {Casertano}, S., {Ferguson}, H. C., {Mobasher}, B., {Gold}, B. {et~al.} 2007, \apj, 659(1),
  98--121.

\bibitem[{{Sullivan} {et~al.}(2011){Sullivan}, {Guy}, {Conley}, {Regnault},
  {Astier}, {Balland}, {Basa}, {Carlberg}, {Fouchez}, {Hardin}, {Hook},
  {Howell}, {Pain}, {Palanque-Delabrouille}, {Perrett}, {Pritchet}, {Rich},
  {Ruhlmann-Kleider}, {Balam}, {Baumont}, {Ellis}, {Fabbro}, {Fakhouri},
  {Fourmanoit}, {Gonz{\'a}lez-Gait{\'a}n}, {Graham}, {Hudson}, {Hsiao},
  {Kronborg}, {Lidman}, {Mourao}, {Neill}, {Perlmutter}, {Ripoche}, {Suzuki},
  \& {Walker}}]{Sullivan11}
{Sullivan}, M., {Guy}, J., {Conley}, A., {Regnault}, N., {Astier}, P., {Balland}, C. {et~al.} 2011, \apj, 737(2), 102.

\bibitem[{{Suzuki} {et~al.}(2012){Suzuki}, {Rubin}, {Lidman}, {Aldering},
  {Amanullah}, {Barbary}, {Barrientos}, {Botyanszki}, {Brodwin}, {Connolly},
  {Dawson}, {Dey}, {Doi}, {Donahue}, {Deustua}, {Eisenhardt}, {Ellingson},
  {Faccioli}, {Fadeyev}, {Fakhouri}, {Fruchter}, {Gilbank}, {Gladders},
  {Goldhaber}, {Gonzalez}, {Goobar}, {Gude}, {Hattori}, {Hoekstra}, {Hsiao},
  {Huang}, {Ihara}, {Jee}, {Johnston}, {Kashikawa}, {Koester}, {Konishi},
  {Kowalski}, {Linder}, {Lubin}, {Melbourne}, {Meyers}, {Morokuma}, {Munshi},
  {Mullis}, {Oda}, {Panagia}, {Perlmutter}, {Postman}, {Pritchard}, {Rhodes},
  {Ripoche}, {Rosati}, {Schlegel}, {Spadafora}, {Stanford}, {Stanishev},
  {Stern}, {Strovink}, {Takanashi}, {Tokita}, {Wagner}, {Wang}, {Yasuda},
  {Yee}, \& {Supernova Cosmology Project}}]{Suzuki12}
{Suzuki}, N., {Rubin}, D., {Lidman}, C., {Aldering}, G., {Amanullah}, R., {Barbary}, K.{et~al.} 2012, \apj, 746(1), 85.

\bibitem[{{Tripp}(1998)}]{Tripp98}
{Tripp}, R. 1998, \aap, 331, 815--820.

\bibitem[{{Wang} \& {Wang}(2013)}]{Wang13}
{Wang}, S., \& {Wang}, Y. 2013, \prd, 88(4), 043511

\end{thebibliography}

\end{document}